\documentclass[12pt]{iopart}
\usepackage{iopams}
\usepackage{graphicx}
\usepackage{epsfig}

\begin{document}

\title{Fluid-fluid versus fluid-solid 
demixing in mixtures of parallel hard hypercubes}

\author{Luis Lafuente$^1$ and Yuri Mart\'{\i}nez-Rat\'on$^2$}

\address{$^1$ Departamento de Matem\'aticas, Escuela Superior 
de Ingenier\'{\i}a, 
Universidad de C\'adiz, Calle Doctor Mara\~n\'on 3, 11002 C\'adiz, Spain.}
\ead{luis.molinero@uca.es}

\address{$^2$ Grupo Interdisciplinar de Sistemas Complejos (GISC), 
Departamento de Matem\'aticas, Escuela Polit\'ecnica Superior, 
Universidad Carlos III de Madrid, Avenida de la Universidad 30, 
28911--Legan\'es, Madrid, Spain.}
\ead{yuri@math.uc3m.es}

\begin{abstract}
It is well known that increase of the spatial dimensionality enhances
the fluid-fluid demixing of a binary mixture of hard hyperspheres,
i.e.\ the demixing occurs for lower mixture size asymmetry
as compared to the three-dimensional case. However, according to simulations,
in the latter dimension the fluid-fluid demixing is metastable with respect to the
fluid-solid transition. According to the results obtained from approximations
to the equation of state of hard hyperspheres in higher dimensions,
the fluid-fluid demixing might become stable for high enough dimension. However, this
conclusion is rather speculative since none of these works have taken into account
the stability of the crystalline phase (by a minimization of a given density functional,
by spinodal calculations or by MC simulations). Of course, the lack of results is justified
by the difficulty of performing density functional calculations or simulations in high dimensions
and, in particular, for highly asymmetric binary mixtures.
In the present work, we will take advantage of a well tested theoretical tool, namely
the fundamental measure density functional theory for parallel hard hypercubes (in the
continuum and in the hypercubic lattice). With this, we have calculated the fluid-fluid and fluid-solid spinodals
for different spatial dimensions. We have obtained, no matter what the dimensionality, the mixture size
asymmetry or the polydispersity (included as a bimodal distribution function centered around
the asymmetric edge-lengths), that the fluid-fluid critical point is always located above the fluid-solid spinodal.
In conclusion, these results point to the existence of demixing between at least one solid phase rich in
large particles and one fluid phase rich in small ones, preempting a fluid-fluid demixing, independently
of the spatial dimension or the polydispersity.
\end{abstract}


\maketitle

\section{Introduction}
\label{introduction}

Entropy-driven demixing was an active line of research since the 1990s. 
Several works had shown the presence of 
fluid-fluid (F-F) demixing 
in binary mixtures 
of hard anisotropic particles \cite{Roij,Wensink,Dubois,Varga,Schmidt1,Martinez-Raton3}. 
The demixed phases can have different  
symmetries, with or without long range 
orientational order, the so called nematic or isotropic phases respectively.  
However, the simplest binary mixture of hard particles 
one can 
imagine, namely the hard sphere (HS) mixture, does not demix into two fluid phases 
even for high mixture size asymmetry. The first simulation study on the fluid-solid
phase separation in the binary mixture of hard spheres \cite{Kranendonk} does not report
any F-F demixing. However, they only consider mixtures with diameter ratios very close to one.
More recent theoretical and simulation works 
on this system consider more asymmetric mixtures and for large
enough asymmetry they have shown that the F-F demixing is metastable 
with respect to the 
fluid-solid (F-S) phase separation, with the crystalline 
phase rich in large spheres while 
the fluid phase is mostly populated by small spheres \cite{Dijkstra,Almarza}.
In these works the F-F instability is taken into account in the computer simulations
by approximating the effective hamiltonian through an explicit depletion potential approximation
(integrating out the degrees of freedom of the small spheres).
The 
same scenario occurs in  binary mixtures of parallel hard cubes 
(PHC) with particles freely moving 
over space  \cite{Martinez-Raton1} or 
constrained on a lattice \cite{Lafuente}. Note that although the cubes are 
anisotropic particles, the gain in the entropic volume after demixing 
can only be reached by changing the mixture composition or the density. As the cubes are parallel, the 
orientational entropy does not play any role here. In this sense the mixture 
of PHC is similar to the HS mixture. However, the effective depletion 
pair potential 
between two large solute particles immersed in a solvent of small particles 
is greater if the particles have cubic symmetry.  
As a matter of fact, this is
why the scaled particle theory (SPT) applied to cubes predicts a 
metastable F-F demixing \cite{Martinez-Raton1},
while the same theory applied to HS does not. 

In contrast to mixtures of additive HS, mixtures of non-additive
HS can exhibit a stable F-F demixing as was shown by several
authors both theoretically and by simulations \cite{noaditivo}.
This behavior can be perfectly understood by resorting to the following
argument: when the interactions between different species are more
repulsive (with a larger excluded volume) than those corresponding to
the same species, the particles in the
demixed configurations have a lower entropic volume and thus a greater
configurational entropy (and consequently a lower free-energy).

The study of the 
hard hypersphere (HHS) one-component fluid has received great  
attention in recent years because certain questions that are ambiguous
in 3D, e.g., jamming, crystallization, glass formation, find clearer
answers when examined from a dimensional perspective. Also the
poor knowledge of the lattice packing properties of HHS in
high dimensions has led to works toward that direction \cite{Cohn}.
Recent works have 
shown that the freezing of HHS is frustrated 
because configurations 
of spheres with simplex liquid ordering are very
different from those corresponding to the 
periodic lattice geometry \cite{Frenkel}. Also calculations 
of the pair correlation functions $g_2(r)$ evince that the short-range 
ordering decreases appreciably with dimensions \cite{Torquato}.
The frustration translates into higher free-energy barriers 
between the fluid and crystal states \cite{Frenkel,Torquato} and  
thus the fluid has a greater propensity (than in three dimensions)
to form a glass upon compression \cite{Torquato}.
Both, simulations 
\cite{Frenkel,Torquato,Estrada} and density functional calculations 
\cite{Finken} have shown that the F-S transition occurs 
at lower packing fractions as the dimensionality increases and also 
that the character of this transition remains of first order. This 
does not mean that freezing is favoured when the dimensionality 
increases, since
close packing is also reached at lower densities. Finally, the HHS fluid 
has been used as a test for different theoretical approximations of 
the structural properties of the fluid, comparing the direct correlation 
function $c(r)$ and the radial distribution function $g(r)$ obtained 
from these theories with those obtained by MC and MD computer simulations 
\cite{Haro,Bishop,Santos}.    

Returning to the demixing behavior of binary mixtures composed 
of particles interacting with hard-core potentials, we should say that 
the number of theoretical studies carried out 
in dimensions higher than three are 
very small, to our knowledge there is only one \cite{Yuste}. 
In this work the authors have shown that the F-F demixing of a binary 
mixture of HHS is favored as the dimensionality increases, i.e. the
demixing shows up at lower mixture size asymmetry when the dimensionality
increases. Notwithstanding, this is not conclusive enough to discard
that one of the demixed phases could be unstable with respect to the
freezing transition. The only way to settle this question is to 
involve the crystalline phase of a binary mixture 
in the theoretical or simulation studies, which obviously
becomes a much more difficult task as compared to three dimensions.

The aim of the present work is to get some insights into what is the most likely 
scenario: F-F versus F-S demixing in a mixture of parallel hard 
hypercubes (PHHC). We have selected 
this system because the freezing of PHHC is much less affected by the geometric frustration 
present in HHS and thus the study is easier to carry out. For 
example, the simple hypercubic lattice is probably the most stable  
configuration of particles in a crystalline phase, and the same should 
occur in the crystalline phase of a binary 
mixture when small cubes occupy a very small fraction 
of the total volume. The other reason behind this selection is related to  
the theoretical tool we will apply to this study: the fundamental measure 
density functional theory developed for a mixture of PHHC defined  
on a lattice \cite{Luis1} or on the continuum \cite{Cuesta}. As has 
already shown, the fundamental measure theory (FMT) 
applied to a HS fluid gives the most accurate results for the freezing 
transition as compared to other density functional theories \cite{Tarazona}. 
Comparison between theory and simulations also confirms the 
high performance of this theory in predicting the equation of state 
of the crystalline phase of PHC \cite{Martinez-Raton1}. Also, we have 
already shown that in three dimensions the F-F demixing is metastable with 
respect to the F-S demixing even for high mixture size asymmetry 
\cite{Martinez-Raton1}. Simulations carried out on binary mixtures of parallel
hard squares found strong clustering of large squares with
crystalline structure in coexistence with a fluid of small squares
\cite{Buhot} which would confirm this scenario in small dimensions. 
Note that this is in accordance with SPT prediction for this system of a stable mixed phase
with respect to the F-F demixing.

In the present paper 
we use FMT for PHHC to perform a F-S bifurcation analysis 
which together with the F-F spinodal 
calculations allow us to conclude that the F-F demixing is preempted 
by the freezing of at least one of the segregated phases (that is rich in large hypercubes). 
We show that this result holds up to the 9th dimension. For dimensions greater than 9 
the third virial 
approximation is accurate enough and we can use it 
to show that the same scenario remains up to dimension 25.
For higher dimensions, we have carried out an asymptotic analysis
of the second virial approach (which becomes exact in the limit of infinite dimension)
and the conclusions do not change: the F-S demixing preempts the F-F demixing. 
Finally, we have included polydispersity in the edge length of the PHHC by means of 
a bimodal distribution function and again we find the same scenario.

The paper is organized as follows. In Sec. \ref{model} we present
the expressions for the direct correlation 
functions of a mixture of PHHC as obtained from the 
FMT defined on a lattice and on the continuum. These functions allow us 
to perform an F-S bifurcation analysis and to calculate the F-F 
demixing spinodals. In Sec. \ref{results} we show the results from these 
calculations. This section is divided into three parts: in Sec. 
\ref{one-component} we present the results from the F-S bifurcation 
analysis of the one-component fluid of PHHC as a function of dimensionality; 
Sec. \ref{demixing} is devoted to a detailed study of the demixing 
transitions as a function of mixture size asymmetry and dimensionality; 
and finally in Sec. \ref{polydispersity} we 
present the results from the bifurcation analysis 
of a bimodal polydisperse mixture 
of PHHC. In Sec. \ref{asymptotic} we carry out an asymptotic analysis 
(for large dimensions) of the second virial approximation which allows us to conclude 
that at infinite dimension the F-F demixing is metastable with 
respect to the F-S demixing. The conclusions are drawn in Sec. \ref{conclusions}
and theoretical details of the bifurcation analysis are relegated to an
appendix. 

\section{Model}
\label{model}
The model consists of a binary mixture of PHHC embedded in dimension $n$ with species $i=1,2$ 
having an edge length equal to $\sigma_i$. We define mixture size asymmetry by 
the coefficient $\kappa=\sigma_2/\sigma_1\geq 1$. The mixture is fully characterized by the density 
profiles $\rho_i({\bf r})$ where the vector 
${\bf r}\equiv (x_1,x_2,\dots,x_n)$ denotes the position of the center of mass of a particle 
in the $n$-dimensional space. If the particle positions are constrained to be on a simple hypercubic lattice the 
variables $x_i$ are set equal to $i\in\mathbb{Z}$ (i.e. we are using the lattice mesh size as the unit of length).
Note that when the lattice mesh size goes to zero, the lattice model goes to the continuum model, therefore
in the lattice model we have chosen $\sigma_2=2$ in order to emphasize its discrete nature.
We will use the density functional (DF) based on the FMT obtained
by us already on a lattice \cite{Luis1} and on the continuum \cite{Cuesta}.  
We refer the reader to those references for a detailed description of the way
to obtain the DF from the free-energy density of a so called zero-dimensional cavity, which
can accommodate one particle at the most. For a more general reference on the theory see Refs.~\cite{Luis2,Tarazona2}.

From the DF we can obtain the Fourier transform of the 
direct correlation functions which are the only expressions we need to compute the F-F
and F-S spinodals corresponding to the F-F or F-S demixing transitions. From these, we can 
analyze the relative stability between F-F and F-S transitions.

\subsection{Direct correlation functions following the FMT 
for PHHC }
\label{dcf}

As we have already shown in \cite{Luis1,Cuesta} the excess part of the free-energy DF corresponding 
to a mixture of PHHC in dimension $n$, $\beta\mathcal{F}_\mathrm{ex}[\{\rho_i\}]$, can be obtained from
the excess free energy of a zero-dimensional cavity
\begin{equation}
\Phi_0(\eta)=\eta+(1-\eta)\ln(1-\eta),
\end{equation}
$\eta$ being the packing fraction of the cavity.

In particular, for the continuum $c$-component mixture of parallel hard hyperparallelepipeds we have
\begin{eqnarray}
\beta\mathcal{F}_\mathrm{ex}[\{\rho_i\}]=\int\mathrm{d}\mathbf{r}\,\Phi_n(\mathbf{r}),\nonumber\\
\Phi_n(\mathbf{r})\equiv\left(\prod_{l=1}^n\sum_{i=1}^{c}\frac{\partial}{\partial\sigma_i^{(l)}}\right) \Phi_0\bigl(\eta(\mathbf{r})\bigr),\label{dimensionCont} \\
\eta(\mathbf{r})=\sum_{i=1}^c 
\int_{x_1-\sigma_i^{(1)}/2}^{x_1+\sigma_i^{(1)}/2}\mathrm{d}x'_1\dots
\int_{x_n-\sigma_i^{(n)}/2}^{x_n+\sigma_i^{(n)}/2}\mathrm{d}x'_n\,\rho_i(\mathbf{r'}),
\end{eqnarray}
$\eta(\mathbf{r})$ being the local packing fraction of a maximal zero-dimensional cavity centered at $\mathbf{r}$ and $\sigma_i^{(l)}$ being the edge length
of component $i$ in the $l$ spatial direction.
The analogous expressions for the discrete case are
\begin{eqnarray}
\beta\mathcal{F}_\mathrm{ex}[\{\rho_i\}]=\sum_{\mathbf{r}\in\mathbb{Z}^n}\Phi_n(\mathbf{r}),\nonumber\\
\Phi_n(\mathbf{r})\equiv\left(\prod_{l=1}^n\Delta_{k_l}\right)\Phi_0\bigl(\eta^{(\mathbf{k})}(\mathbf{r})\bigr),\qquad\Delta_k f(k)\equiv f(1)-f(0), \label{dimensionLat}\\
\eta^{(\mathbf{k})}(\mathbf{r})=\sum_{i=1}^c 
\sum_{x'_1=x_1-\sigma_i^{(1)}/2+1-k_1}^{x_1+\sigma_i^{(1)}/2-1}\dots
\sum_{x'_n=x_n-\sigma_i^{(n)}/2+1-k_n}^{x_n+\sigma_i^{(n)}/2-1}\rho_i(\mathbf{r'}),\quad\mathbf{k}\in\{0,1\}^n, \label{etak}
\end{eqnarray}
$\eta^{(\mathbf{k})}(\mathbf{r})$ being the packing fraction of
a zero-dimensional cavity around $\mathbf{r}$. For a connection between the
difference operator in equation (\ref{dimensionLat}) and the M\"obius inversion formula, the reader is referred to \cite{Luis2}.

Using Eqs. (\ref{dimensionCont}) and (\ref{dimensionLat}), we can calculate the direct correlation 
functions defined through the second 
functional derivative,
$\displaystyle{c_{ij}^{(n)}({\bf r}_{12})=-\delta^2\beta {\cal F}_{\rm ex}^{(n)}/\delta\rho_i({\bf r}_1)
\delta\rho_j({\bf r}_2)}$, 
evaluated at uniform density profiles $\rho_i({\bf r})=\rho_i$. If we 
calculate the Fourier 
transforms of these functions, evaluate them at the wave vector 
${\bf q}=q(1,0,\dots,0)$ and take $\sigma_i^{(k)}=\sigma_i$ 
for any $k$, in the continuum we obtain
\begin{eqnarray}
\fl
-\hat{c}_{ij}^{(n)}(q)=
\sum_{k=0}^{n-1}
C^{n-1}_k(\sigma_i\sigma_j)^k\sigma_{ij}^{n-1-k}\nonumber\\
\times
\left[\zeta_k\sigma_{ij}j_1(q\sigma_{ij}/2)+
\zeta_{k+1}\sigma_i\sigma_jj_1(q\sigma_i/2)j_1(q\sigma_j/2)\right],
\label{note}
\end{eqnarray} 
with $\sigma_{ij}=\sigma_i+\sigma_j$, $C^n_k=n!/k!/(n-k)!$
the usual combinatorial coefficients and $j_1(x)$ the spherical 
Bessel function of first order. The functions 
$\zeta_k(\xi_{n-k},\dots,\xi_n)$ with $k=0,\dots,n$ 
can be obtained recurrently from 
$\zeta_0(\xi_n)=1/(1-\xi_n)$ 
using the following relations
\begin{equation}
\zeta_{k+1}(\xi_{n-k-1},\dots,\xi_n)
=\sum_{i=0}^k\xi_{n-i-1}\frac{\partial \zeta_k(\xi_{n-k},\dots,\xi_n)}
{\partial \xi_{n-i}}, \quad k=0,\dots, n-1 
\label{recurre}
\end{equation}
where we have defined $\xi_k=\sum_i \rho_i\sigma_i^k$.   
Note that the packing fraction $\eta$ is just equal to $\xi_n$. 
For the lattice version we have
\begin{eqnarray}
\fl
-\hat{c}_{ij}^{(n)}(q)=\sum_{k=0}^{n-1}C_k^{n-1}(-1)^{n-k}
(\sigma_i\sigma_j)^k\left[(\sigma_i-1)(\sigma_j-1)\right]^{n-k-1}
\frac{1}{\sin^2(q/2)}\nonumber\\
\times\left[\frac{\sin(q\sigma_i/2)\sin(q\sigma_j/2)}{1-\eta_{k+1}}
-\frac{\sin(q(\sigma_i-1)/2)\sin(q(\sigma_j-1)/2)}{1-\eta_k}\right],
\label{lattice}
\end{eqnarray}
where we have defined $\eta_k=\sum_i\rho_i\sigma_i^k(\sigma_i-1)^{n-k}$ 
for $k=0,1,\dots,n$, which is the uniform version of $\eta^{(\mathbf{k})}(\mathbf{r})$ with
$k\equiv k_1+\cdots+k_n$.
 
The fluid pressure of the mixture in the continuum 
is related to the function $\zeta_n(\xi_0,\dots,\xi_n)$ 
through the relation $\zeta_n=\partial \beta p/\partial \xi_n$. 
The lattice version give us the following expression for the pressure
\begin{equation}
\beta p=-\sum_{k=0}^n C_k^n(-1)^{n-k}\ln(1-\eta_k).
\end{equation}

We should note that the expressions (\ref{note}) and (\ref{lattice}) 
in the low density limit 
$\rho_i\to 0$ recover the exact
second and third virial approximations. 

\subsection{F-F and F-S spinodals}
\label{f-f_f-s}

The packing fraction $\eta$ (note that $\eta=\xi_n$ in the continuum and $\eta=\eta_n$ in the lattice)
and the wave number $q$ at the 
F-S transition of the binary mixture can be calculated from the singularity 
of the matrix with coefficients defined through the partial structure factors 
\begin{equation}
S_{ij}(q,\eta)=\delta_{ij}-\rho\sqrt{x_ix_j}\hat{c}_{ij}^{(n)}(q),
\end{equation}
with $x_i=\rho_i/\rho$ the mixture composition and $\rho=\sum_i\rho_i$ 
the total density. The singularity condition gives us, at a fixed composition $x_1$ ($x_2=1-x_1$), 
\begin{eqnarray}
\Delta(q,\eta,x_1)&\equiv& \det
\left(
\begin{array}{cc}
1- \rho x_1\hat{c}^{(n)}_{11}(q) & -\rho\sqrt{x_1x_2}\hat{c}^{(n)}_{12}(q)\\
-\rho\sqrt{x_1x_2}\hat{c}^{(n)}_{12}(q) & 1-\rho x_2 \hat{c}^{(n)}_{22}(q)
\end{array}
\right)
\nonumber \\
&=&\left(1-\rho_1\hat{c}^{(n)}_{11}(q)\right)\left(1-\rho_2\hat{c}^{(n)}_{22}(q)\right)
-\rho_1\rho_2\left(\hat{c}_{12}^{(n)}(q)\right)^2=0.
\end{eqnarray}
Now, we should look for the lowest value of $\eta$ that satisfied this equation.
If we take into account that $\Delta(\eta,q,x_1)=0$ defines a function $\eta_\mathrm{sing}(q,x_1)$,
in order to find the values $\eta_\mathrm{bif}(x_1)$ and $q_\mathrm{bif}$ at the F-S transition we need to solve
\begin{equation}
\eqalign{
\eta_\mathrm{bif}(x_1) = \min_{q> 0} \eta_\mathrm{sing}(q,x_1),\cr
q_\mathrm{bif} = \arg \min_{q>0} \eta_\mathrm{sing}(q,x_1).
}
\label{FSspinodal}
\end{equation}
In the continuum this is equivalent to the set of equations
\begin{equation}
\Delta(q,\eta,x_1)=\frac{\partial \Delta(q,\eta,x_1)}{\partial q}=0,
\label{lasec}
\end{equation}
In the lattice version, due to the discrete nature of $q$ (in units of the lattice mesh size) we
do not have the second equality in (\ref{lasec}). Finally, in both cases the F-F spinodal can be calculated
from the condition 
$\Delta(0,\eta,x_1)=0$.

\section{Results}
\label{results}

This section is divided into three parts. The first one is devoted to showing
the performance of our theory concerning the calculation of the F-S
transition of a one-component fluid of PHHC as a function of dimensionality
$n$. In the second part we analyze the relative stability between the F-F
and F-S demixing, again as a function of $n$, of (i) a binary mixture of
PHHC and of (ii) a bimodal polydisperse fluid. For the former system, we have
carried out the analysis both in the continuum and in the lattice cases,
while for the latter only the continuum system is studied.
The last part of the section is devoted to the study of the binary mixture phase behavior
at infinite dimension, by means of an asymptotic analysis.

\subsection{One component fluid}
\label{one-component}
As we have already shown using both versions of the FMT,
the continuum parallel hard cube fluid in dimension three exhibits
a second order F-S transition  where the cubes crystallize in a simple
cubic lattice \cite{Cuesta}; in the lattice version \cite{Luis1}  the system
shows a second order transition where the structure of the ordered phase
depends on the size of the cubes (from smectic for the smallest cubes possible
to solid for larger sizes). Herein we will be referring to F-S demixing for both
the continuum and lattice model, but we ask the reader to take into account that in the
discrete case the structure of the ordered phase could be different from a solid.
In the continuum the same behavior was also predicted
from simulations \cite{Jagla,Bela,Hoover2}. Although the character (second order
vs. first order), and the symmetry of the crystalline phase were
adequately predicted from FMT as compared to MC simulation results,
the precise location of the transition point is underestimated by the
theory \cite{Martinez-Raton1}. This discrepancy is explained
by noting that the uniform limit of the FMT recovers the SPT and, therefore,
the fluid phase predicted by FMT inherits all its defects. While
all the virial coefficients predicted by SPT are positive, it is known
that this is not the real case \cite{Hoover1,Swol} where the presence of negative
coefficients results in a poor convergence of the virial expansions. As a consequence,
although FMT recovers the low-density limit, the equation of state for the fluid
near the transition point is not very accurate.
Notwithstanding, because of the dimensional crossover property of the FMT DFs,
the predictive power of FMT is greatly enhanced when the packing fraction of
the crystalline phase is high enough. Actually, for packing fractions
$\eta \geq 0.5$ the comparison between theory and simulations is fairly good \cite{Martinez-Raton1}.
Finally, for dimensions higher than three we should remark that there is a
rigorous result that shows that at infinite dimension the PHHC fluid exhibits a second order phase transition to
a crystalline phase with a simple hypercubic symmetry \cite{Kirkpatrick}.

\begin{figure}
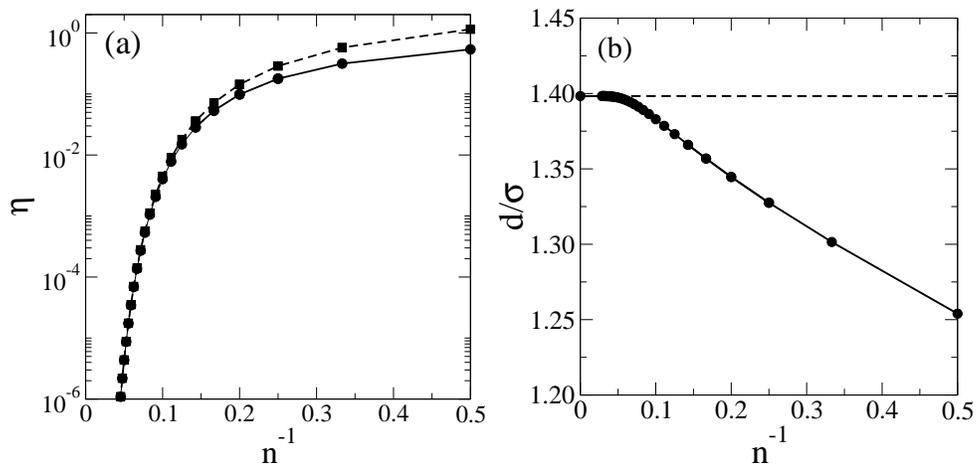

\hspace*{1.5cm}
\epsfig{file=fig1a.eps,width=2.5in}
\epsfig{file=fig1b.eps,width=2.5in}
\caption{Packing fraction $\eta$ (a) and period of the crystalline 
phase $d/\sigma$ (b) at bifurcation as a function of the dimensionality
$n$.}
\label{fig0}
\end{figure}

We calculate here the F-S transition of the continuum PHHC as a function of
the dimensionality using a bifurcation analysis, i.e.\ by solving
the equations (\ref{lasec}) in the one-component limit 
to find the packing fraction $\eta_\mathrm{bif}$ and the
period in units of the edge length $d/\sigma$ at the transition
point (note that $d$ is related to $q_\mathrm{bif}$ as $d/\sigma=2\pi/q_\mathrm{bif}$).
These values are plotted in figures \ref{fig0} (a) and (b),
respectively. As is already known, the second virial approximation
becomes exact in the limit $n\to \infty$. Here, the packing
fraction at the transition scales with $n$ as $\eta\sim 2^{-n}$.
With the aim of comparison, we have also plotted in figure \ref{fig0} the
results from this approach. As
we can see from the figure, the packing fraction $\eta_\mathrm{bif}$ and the period $d/\sigma$
obtained from both FMT and second virial approximations are very similar when
$n\sim 10$. Therefore, it should be expected that around this value of the dimension
and for higher ones FMT predictions should not differ too much from those obtained
by means of a virial expansion. This supports our approach in the next subsections where we have analyzed
the binary mixture stability using FMT only up to $n=9$ and from there to $n=25$ we
have used the third virial expansion instead, and we have concluded with an asymptotic analysis of
the second virial expansion, which should be accurate enough for higher dimensions.

\subsection{F-F versus F-S demixing in binary mixtures of PHHC}
\label{demixing}
\begin{figure}
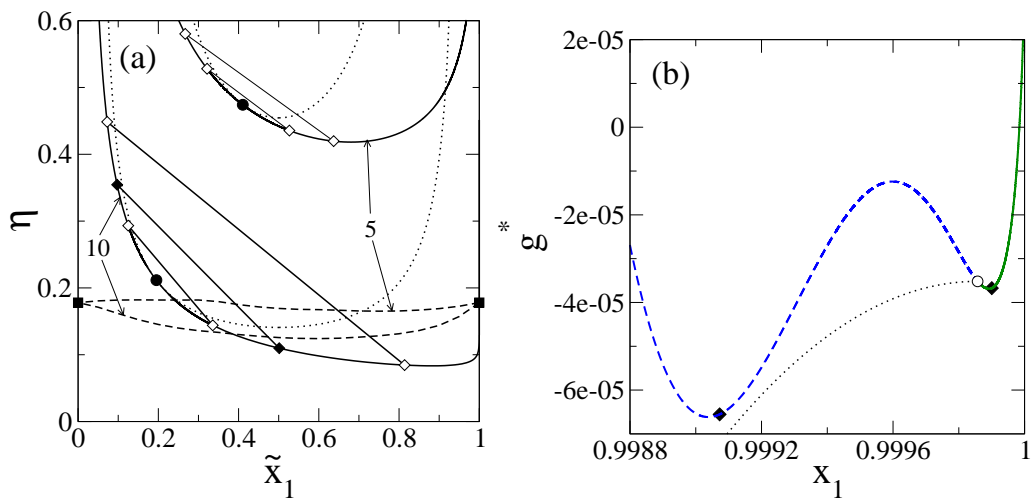

\begin{center}
\epsfig{file=fig2a.eps,width=2.5in}
\epsfig{file=fig2b.eps,width=2.8in}
\caption{\label{rr}(a) F-F (dotted) and F-S (dashed) spinodals in the packing fraction-
composition plane for a binary mixture of PHHC in dimension four 
and mixture asymmetries equal to $\kappa=5$ and $\kappa=10$.  
The coexistence F-F binodals are also plotted with 
solid lines. The solid circles indicate the position of the critical 
points while the diamonds joined by straight lines indicate  
some examples of coexisting fluid phases (note that one or both of them 
are metastable with respect to the freezing transition). The solid 
squares indicate the F-S transition of the pure one-component fluid. 
(b) The Gibbs free energy per particle in reduced units minus a straight line 
(to make the coexistence more visible) 
$g^*=\beta G/N-7.4962-1273.2x_1$ as a function of the molar 
fraction of the small 
component. The mixture size asymmetry is
$\kappa=10$ and the pressure is fixed to 
$\beta p\sigma_1^3=939.49$ [coinciding with that corresponding to the 
coexistence points labeled with filled diamonds in (a)]. 
The filled diamonds 
indicate the F-F coexisting 
points while the open circle indicates the F-S 
bifurcation point. To the left of this point the fluid branch (with the dashed 
line) is unstable with respect to the solid branch sketched (not calculated) 
with the dotted line. 
}
\end{center}
\end{figure} 

In this section, we will analyze the phase behavior of the binary 
mixture of PHHC as a function of the dimensionality. As was 
already shown in dimension three, the F-F demixing is always 
unstable with respect to the F-S demixing in which a 
crystalline phase preferentially populated by large
cubes coexists with a fluid phase 
rich in small cubes \cite{Martinez-Raton1}. The uniform mixture becomes 
unstable with respect to the F-F demixing for a mixture 
size asymmetry $\kappa = 5 + \sqrt{24} \approx 10$ in the continuum and $\kappa = 13$ in the lattice.
However, the position of the critical  
point of the F-F spinodal curve in the packing fraction-composition plane 
is always located above the F-S spinodal curve, the latter computed from 
the set of equations (\ref{FSspinodal}). This is equivalent to saying that
at least one of the coexisting uniform phases is unstable with respect to the freezing transition. 

\begin{table}
\begin{center}
\begin{tabular}{|c|c|c|c|c|c|c|}
\hline
$n$ & 3 & 4 & 5 & 6 & 7\\
\hline
$\kappa_{\rm{cont}}^*$ & 9.90 & 3.73 & 2.54 & 2.09 & 1.87\\
\hline
$\kappa_{\rm{latt}}^*$ & 13 & 5 & 3 & 3 & 2\\
\hline
\end{tabular}
\caption{\label{tabla} Critical aspect ratio $\kappa^\ast$ for which the F-F demixing appears for the first time in the continuum and lattice models.} 
\end{center}
\end{table}
We have found the same scenario for higher dimensions. The main
differences are only quantitative and can be summarized as follows:
the lowest mixture size asymmetry at which the F-F phase
separation shows up (see table \ref{tabla}) decreases dramatically with the dimensionality and
the same happens with the packing-fraction value at the critical point.
Thus the demixing transition is enhanced with $n$. On the other hand,
the packing-fraction values at the F-S spinodal
also decrease with $n$, and although they get closer to the F-F spinodal
in the neighborhood of the critical point, they remain always below it.
This behavior is present even for high values of the mixture size asymmetry.
It is worth noting that $\kappa^\ast$ for the continuum model decreases as the
inverse of the dimension; actually from the values shown in table \ref{tabla}
we have found that $\kappa^\ast$ follows very accurately the law
$\kappa^\ast=1+a/(n-b)$ where $a=3.809$ and $b=2.566$.
The conclusion is that, no matter what the dimension or the mixture size asymmetry,
the most probable scenario is the presence of 
demixing between a fluid phase rich in small cubes and a crystalline phase mostly populated
by large ones. This scenario is illustrated in figure \ref{rr} where
we have plotted both spinodals (F-F and F-S) and the F-F coexistence binodals
obtained from the continuous model 
of PHHC in dimension four and asymmetries fixed to
$\kappa=5$ and $\kappa=10$. Note that we have used to represent
the composition of the mixture the volume composition $\tilde{x}_1$, which
is related to the molar fraction as $\tilde{x}_i=\sigma_i^n x_i/\langle \sigma^n \rangle$,
where $\langle \sigma^n \rangle \equiv \sum_i \sigma_i^n x_i$.
As we can see from the figure, the critical point in the F-F spinodal
$(\tilde{x}_1^\mathrm{c},\eta^\mathrm{c})$ is located
above the F-S spinodal, therefore at least one of
coexisting fluid phases is
unstable with respect to the freezing transition. This scenario is 
illustrated in figure \ref{rr}(b), where we plot the Gibbs free-energy 
per particle as a function of the molar fraction of small cubes 
$x_1$ for a fixed value of the fluid pressure (note the difference 
between $\tilde{x}_1$ and $x_1$ due to the huge difference in volume
between the large and small cubes). 
From the left of the F-S bifurcation point (open circle), the fluid branch is usually located 
above the solid branch which is sketched with a non-convex curve. 
From our earlier calculations in $n=3$ the convexity of the solid branch 
is usually recovered when 
the volume fraction of big cubes is near unity. As we have already shown in \cite{Martinez-Raton2}
the infinite mixture size asymmetry limit corresponds to a one-component fluid of
sticky cubes. For this case, at any fixed solvent fugaticy, the coexistence is between an infinite dilute vapor
phase and a close-packed crystal. When we include a small amount of polydispersity in
the edge length of the large cubes the convexity of the solid branch is recovered at
high packing fractions, and the coexistence is now between a highly dense crystal
and a vapor phase. We expect a similar behavior for a highly asymmetric mixture.

\begin{figure}
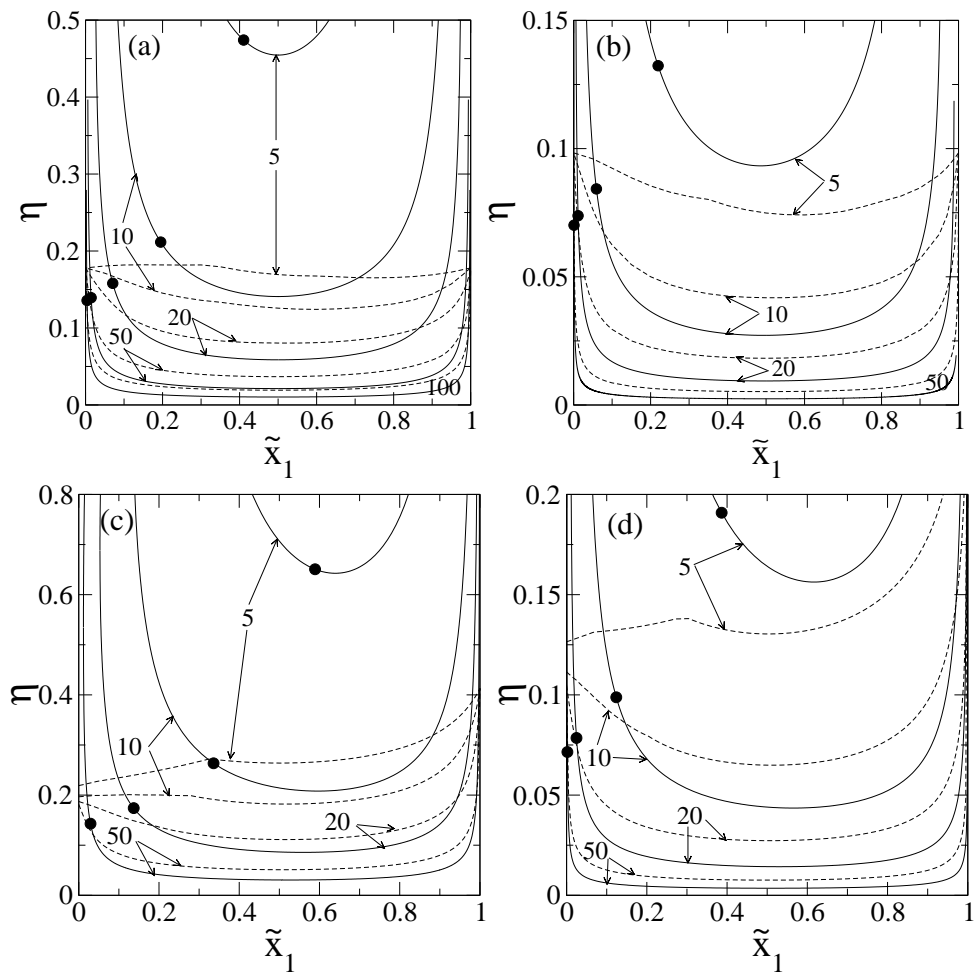

\begin{center}
\epsfig{file=fig3a.eps,width=2.4in}
\epsfig{file=fig3b.eps,width=2.5in}
\epsfig{file=fig3c.eps,width=2.5in}
\epsfig{file=fig3d.eps,width=2.5in}
\end{center}
\caption{Phase behavior of a binary mixture of PHHC in 
the continuum [(a) and (b)] and in the lattice [(c) and (d)] for
dimensions 4 [(a) and (c)] and 
5 [(b) and (d)]. The solid and dashed lines represent the F-F and F-S spinodals 
for different values of the mixture asymmetry $\kappa$ 
(correspondingly labeled in the figure). The filled circles indicate
the locations of the F-F critical points.}
\label{fig1}
\end{figure}

In figure \ref{fig1}, we show the phase behavior of both continuum
(figures (a) and (b)) and lattice (figures (c) and (d))  PHHC 
for $n=4$ and $n=5$ for a large set of mixture size asymmetries, some of 
them being as high as $\kappa=100$. It is remarkable that  
even for very high values of $\kappa$ the behavior is the same as that 
described above: the F-S demixing is the most likely scenario.

\begin{figure}
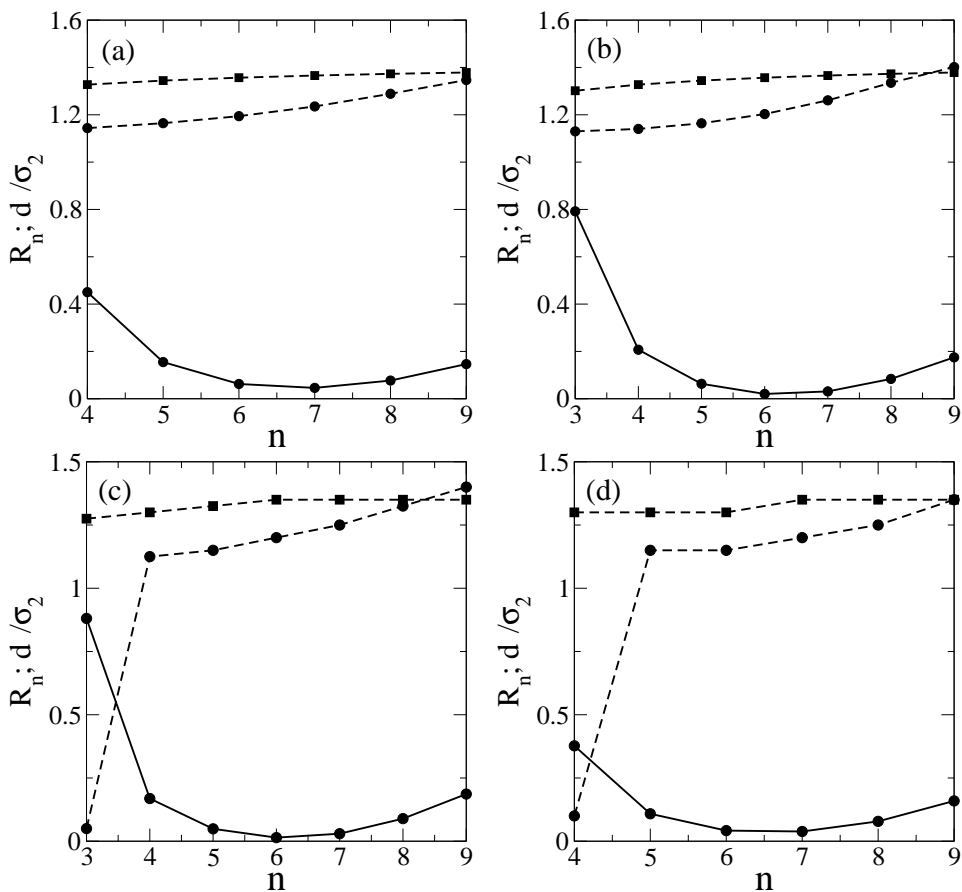

\begin{center}
\epsfig{file=fig4a.eps,width=2.5in}
\epsfig{file=fig4b.eps,width=2.5in}
\epsfig{file=fig4c.eps,width=2.5in}
\epsfig{file=fig4d.eps,width=2.5in}
\caption{The coefficient $R_n$ (solid line) and the  
lattice periods $d/\sigma_2$ corresponding to the crystalline 
mixed phase (the circles joined with dashed line) 
and to the one-component crystal (squares joined with dashed line) 
as a function of $n$ as obtained from the continuum [(a) and (b)] 
and the lattice [(c) and (d)] models. The values for the 
mixture asymmetries are
$\kappa=10$ [(a) and (c)] and $\kappa=20$ [(b) and (d)].} 
\label{fig2}
\end{center}
\end{figure}

This trend can be better appreciated in figures \ref{fig2}(a) and (b) (for the continuous 
model) and (c) and (d) (for the lattice model), where we plot the 
coefficient $R_n\equiv\bigl(\eta^\mathrm{c}-\eta_\mathrm{bif}(x_1^\mathrm{c})\bigr)/\eta_\mathrm{bif}(x_1^\mathrm{c})$ which measures the 
relative distance between both spinodals at the F-F critical point 
as a function of $n$. As we can see from the figure, this coefficient
decreases with $n$ up to a certain dimension (the position of
the minimum) which depends on $\kappa$, and further increases
with $n$ diverging at $n\to \infty$ (this will be shown by means of the
asymptotic analysis presented in section \ref{asymptotic}).
In these figures are also shown the period $d/\sigma_2$ of the crystalline mixed
phase with $\tilde{x}_1=\tilde{x}_1^\mathrm{c}$ 
calculated at bifurcation 
as a function of $n$, and the period of the one-component crystal. 
In general, for low dimensions the crystalline phase of large cubes
becomes more packed 
by adding an small amount of small cubes, an effect clearly related with 
the entropic depletion forces.

\begin{figure}
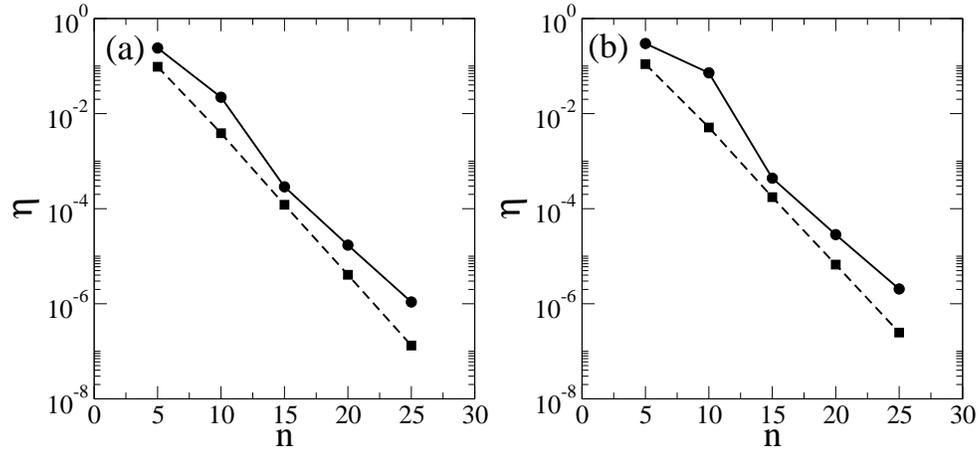

\hspace*{1.5cm}
\epsfig{file=fig5a.eps,width=2.5in}
\epsfig{file=fig5b.eps,width=2.5in}
\caption{The packing fractions corresponding to the the F-F critical point and 
to the the F-S transition at $\tilde{x}_1^\mathrm{c}$ as a function of dimensionality 
following the third virial approximation as obtained from 
the continuum (a) and lattice (b) models.}
\label{figg}
\end{figure}

Using the third virial approximation [equations (\ref{note}) 
and (\ref{lattice}) up to first order in density] we have 
extended the spinodal calculations up to dimension 25. In figure (\ref{figg})
we plot the packing fractions of the F-F critical point
and of the F-S transition at $\tilde{x}_1^\mathrm{c}$ as a function of $n$ in
both the continuum and the lattice models. Again, we find the
expected behavior: the freezing preempts the F-F demixing. 

\subsection{Effect of polydispersity on the F-F and F-S demixing}
\label{polydispersity}

\begin{figure}
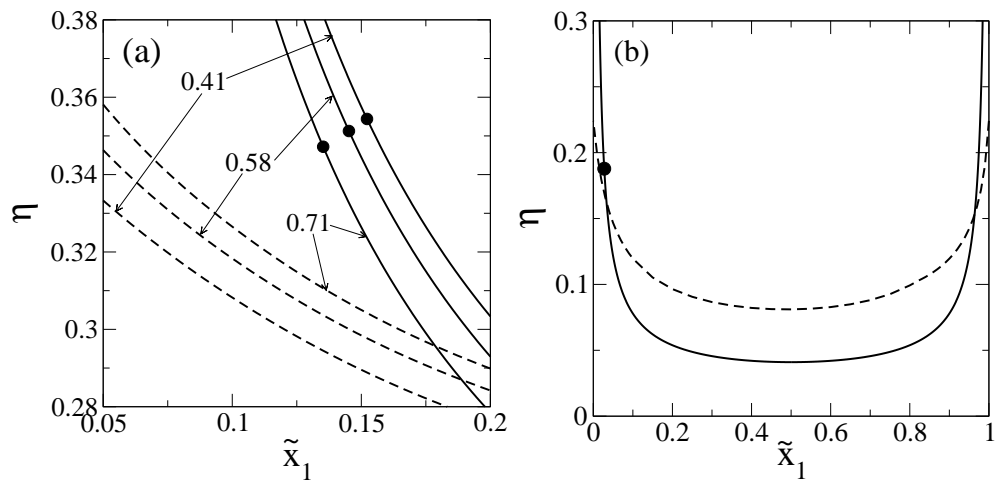

\hspace*{1.5cm}
\epsfig{file=fig6a.eps,width=2.6in}
\epsfig{file=fig6b.eps,width=2.5in}
\caption{(a) F-F and F-S spinodals of PHHC in dimension three 
around the F-F critical point for 
different values of the polydisperse coefficient $r$ as labeled in 
the figure. (b) 
The whole spinodals in dimension four at polydispersity 
$r=0.577$. The mixture size asymmetry is 
fixed to $\kappa=50$ in (a) and $\kappa=20$ in (b). 
The circles specify the position of the critical 
points.}
\label{fff}
\end{figure}

We have also studied the effect that the polydispersity has on the 
demixing behavior of PHHC. We have relegated to the appendix the 
formalism that leads to the main equations for the F-F and F-S spinodals.
If we introduce polydispersity 
in the cube-edge lengths around their mean values 
$\sigma_1=1$ and $\sigma_2=50$ (i.e.\ following a bimodal distribution 
function) for 
$n=3$ and $\sigma_2=20$ for $n=4$, we find 
that the crystalline phase destabilizes with respect to the fluid 
phase as the polydispersity increases. However, although the 
F-F critical point gets closer to the 
F-S spinodal it is still located above it, even for large 
polydispersities (see figures \ref{fff} (a) and (b)).

\section{Asymptotic analysis}
\label{asymptotic}
In this section we will carry out the asymptotic analysis for the functions $\eta^\mathrm{c}(n)$ 
and $\tilde{x}_1^\mathrm{c}(n)$ corresponding to the position of the F-F critical 
point as a function of dimensionality when $n\to\infty$. For this, we will use the second
virial approach, as it becomes exact at this limit. The aim of the analysis is to
find the relative position of the F-F critical point with respect to the F-S transition
at large dimensions. In order to do that, we will use the asymptotic expressions for the
F-F critical point to evaluate $\Delta(q_\mathrm{min},\eta^\mathrm{c},x_1^\mathrm{c})$ (with $q_\mathrm{min}$
the position of the absolute minimum of $\Delta(q,\eta^\mathrm{c},x_1^\mathrm{c})$ as a function of $q$).
Note that if this quantity is negative, this means that the F-S transition is below
the F-F critical point. For the sake of simplicity, we will restrict this analysis to the
continuum model. In the lattice case, the expressions become too cumbersome, but qualitatively there
is no difference with the continuum model, so we have preferred not to include it in this paper.

The expression for the direct correlation functions $\hat{c}_{ij}^{(n)}(q)$ in the
second virial approximation can be obtained from equation (\ref{note}) as a Taylor expansion
of zeroth order in the densities. In this approximation, the pressure can be cast as
\begin{equation}
\beta p=\rho+\frac{1}{2}\sum_{i,j}\rho_i\rho_j\sigma_{ij}^n.
\label{2nd}
\end{equation}

In the following, we will use the volume composition $\tilde{x}_i$
and the scaled packing fraction $\eta_0\equiv\eta 2^n$ as the adequate variables to evaluate
$\beta p$ and $\Delta(q,\eta,x_1)$ when $n\gg1$. 
In these new variables their expressions become
\begin{equation}
\fl
p^\ast(\eta_0,\tilde{x}_1)\equiv \beta p(2\sigma_1)^n=\eta_0\left(\tilde{x}_1+\frac{\tilde{x}_2}{\kappa^n}\right)+
\frac{\eta_0^2}{2}\left[\tilde{x}_1^2+\frac{\tilde{x}_2^2}{\kappa^n}+2\tilde{x}_1\tilde{x}_2
\left(\frac{1+\kappa^{-1}}{2}\right)^n\right],\label{sspin0}
\end{equation}
for the pressure and
\begin{eqnarray*}
\fl
\Delta(q,\eta_0,\tilde{x}_1) = 1+\tilde{x}_1\eta_0j_1(q\sigma_1)+
\tilde{x}_2\eta_0 j_1(q\sigma_2)+
\tilde{x}_1\tilde{x}_2\eta_0^2\\
\times\left[j_1(q\sigma_1)j_1(q\sigma_2)-\left(\frac{\kappa+\kappa^{-1}+2}{4}\right)^n
j_1^2(q\sigma_{12}/2)\right].
\end{eqnarray*}
Note that $\tilde{x}_2=1-\tilde{x}_1$.
The above expression evaluated at $q=0$ gives 
\begin{equation}
\Delta(0,\eta_0,\tilde{x}_1)=1+\eta_0-\tilde{x}_1\tilde{x}_2\delta_n\eta_0^2,
\end{equation}
where we have defined the coefficient $\delta_n\equiv\kappa^{-n}[(\kappa+1)/2]^{2n}-1$. The equation 
$\Delta(0,\eta_0,\tilde{x}_1)=0$ defines the F-F spinodal:
\begin{equation}
\eta_0^\mathrm{FF}(\tilde{x}_1)=\frac{1+\sqrt{1+4\delta_n\tilde{x}_1\tilde{x}_2}}{2\tilde{x}_1\tilde{x}_2\delta_n}.
\label{sspin}
\end{equation}
The critical point can be found as the value of $\tilde{x}_1$ at which
$\mathrm{d}p^*(\eta_0^\mathrm{FF},\tilde{x}_1)/\mathrm{d}\tilde{x}_1=0$. When
$n\to\infty$, the composition at the critical point $\tilde{x}_1^\mathrm{c}$
goes to zero (and therefore $\tilde{x}_2^\mathrm{c}\asymp 1$); taking this into account and
using the new variables $s_n\equiv \sqrt{1+4\delta_n\tilde{x}_1}-1$ and
$\tau_n\equiv \delta_n/\kappa^n$ we have
\begin{equation}
\frac{\mathrm{d}p^*}{\mathrm{d}\tilde{x}_1}\asymp\frac{2+s_n}{2s_n^2(1+s_n)}\left[s_n^3-4\sqrt{\tau_n}(1+\sqrt{\tau_n})s_n-8\tau_n\right],
\qquad(n\to\infty).
\end{equation}
The asymptotic solution of $\mathrm{d}p^*(\eta_0^\mathrm{FF},\tilde{x}_1)/\mathrm{d}\tilde{x}_1=0$ with respect to $s_n$ 
gives us $s_n\asymp \sqrt{\tau_n}+\sqrt{4\sqrt{\tau_n}+\tau_n}$ and then, at the F-F critical point,
\begin{equation}
\tilde{x}_1^\mathrm{c}(n)\asymp(2\delta_n)^{-1}\left[3\sqrt{\tau_n}+\tau_n+(1+\sqrt{\tau_n})
\sqrt{4\sqrt{\tau_n}+\tau_n}\right],
\quad(n\to\infty),
\end{equation}
and from equation (\ref{sspin})
\begin{equation}
\eta_0^\mathrm{c}(n)\asymp\tau_n^{-1/4}\sqrt{1+\frac{\sqrt{\tau_n}}{4}}-\frac{1}{2},
\qquad(n\to\infty).
\end{equation}
From here we have that the leading terms in the asymptotic behavior at the F-F critical point are
$\tilde{x}_1^\mathrm{c}(n)\asymp\left[2\sqrt{2\kappa}/(\kappa+1)^{3/2}\right]^n$ and  
$\eta_0^\mathrm{c}(n)\asymp \left[2/\left(1+\kappa^{-1}\right)\right]^{n/2}$, respectively.
Note that although $\eta_0\to\infty$ when $n\to\infty$ the packing fraction $\eta=2^{-n}\eta_0$ 
goes to zero when $n\to\infty$.

The function $\Delta(q,\eta^\mathrm{c}_0,\tilde{x}_1^\mathrm{c})$ is asymptotically
\begin{equation}
\Delta(q,\eta_0^\mathrm{c},\tilde{x}_1^\mathrm{c})\asymp1-j_1^2(q\sigma_{12}/2)+\eta_0^\mathrm{c}\left[j_1(q\sigma_2)-j_1^2(q\sigma_{12}/2)
\right],
\quad(n\to\infty). 
\label{ss}
\end{equation} 
When $n\to\infty$, $\eta_0^\mathrm{c}\to\infty$, and the dominant contribution to $\Delta(q,\eta_0^\mathrm{c},\tilde{x}_1^\mathrm{c})$ 
is just the term proportional to $\eta_0^\mathrm{c}$ in equation (\ref{ss}) which for $\kappa\gg1$ has 
its absolute minimum at $q_\mathrm{min}\sigma_2=3.506$ with the value
$\Delta(q_\mathrm{min},\eta_0^\mathrm{c},\tilde{x}_1^\mathrm{c})=-0.416\eta_0^\mathrm{c}$. 
This negative value 
implies that the mixture with composition and packing fraction corresponding 
to the F-F critical point is unstable with respect to the F-S transition. Thus, the F-F demixing 
is unstable with respect to the F-S demixing.

\section{Conclusions}
\label{conclusions}
In the present work we have made the first attempt to answer to the question 
about how the dimensionality and mixture size asymmetry affect
the relative stability of the F-F demixing of PHHC against the F-S phase 
separation, based on the bifurcation analysis for the latter transition. 
Although the results presented here strongly suggest the F-S demixing as the 
most likely scenario independently of the dimensionality and of the mixture 
size asymmetry, the final confirmation 
of this phase behavior 
can be given only after carrying out MC simulations or DFT coexistence 
calculations on this system. Both tasks  
become much more difficult as the mixture size asymmetry and 
dimensionality increase.
The main difficulty in performing MC simulations consists of the high probability of overlapping 
many small cubes when a given large cube is moved by a MC step.
Thus a more sophisticated simulation scheme like a cluster movement algorithm is required to carry 
out these simulations.
Even using these schemes, the increase of spatial dimensionality (apart from the high mixture asymmetry) 
constitutes an additional complication due to the huge increase in the  number of particles
necessary to perform a simulation with a reliable statistics. Instead of performing an MC 
simulation on the binary mixture, it is possible to map the mixture onto an effective one-component fluid 
with particles interacting via an effective depletion potential which can be previously calculated using 
the formalism described in \cite{Dijkstra}. Although this mapping is approximate it should 
work very well for very asymmetric mixtures.

The extrapolation of this result to the HHS mixture or other additive hard mixtures
should be made with some caution. The elevated barrier for freezing
in higher dimensions might allow for the metastable F-F separation to be observed
and particularly long lived. Also, as we have pointed out in section \ref{introduction}, the 
freezing transition in HHS becomes more frustrated with dimensionality. 
Thus it might possibly occur that at some dimension 
this frustration makes the F-F 
demixing the stable one. To elucidate this question we should wait 
for future works which we hope will have been motivated by the present one. 

\section*{Acknowledgements}

We acknowledge support from the Direcci\'on
General de Universidades e Investigaci\'on of the Comunidad de
Madrid (Spain), under the R\&D Programmes of activities 
MODELICO-CM/S2009ESP-1691 
and to the Ministerio de Educaci\'on y Ciencia of Spain under the grants
MOSAICO and FIS2010-22047-C05-C04

\appendix
\section{Polydisperse PHHC}
\label{a_poly}

In this appendix, we present the formalism used to prove that the transition to the crystalline phase 
preempts the F-F 
demixing even in polydisperse mixtures of PHHC. 
For this purpose we select a 
bimodal length distribution function as 
\begin{equation}
h(\sigma)=xh_0(\sigma/\sigma_1)+(1-x)h_0(\sigma/\sigma_2),
\label{cero}
\end{equation}
where $0\leq x\leq 1$ is the polydisperse composition variable, and 
we use for $h_0(\sigma/\sigma_i)$ the normalized to unity 
Schultz distribution function
with mean values fixed to $\sigma_i$ ($i=1,2)$:
\begin{equation} 
h_0(\sigma/\sigma_i)=\frac{(\nu+1)^{\nu+1}}
{\sigma_i\Gamma(\nu+1)}\left(\frac{\sigma}{\sigma_i}\right)^{\nu}
\exp\left[-(\nu+1)\sigma/\sigma_i\right],
\end{equation}
where $\Gamma(x)$ is the Gamma function and $\nu$ controls the width 
of both peaks located in the neighborhood of $\sigma_i$ and
related with the polydisperse coefficient $r$ of the distribution 
function $h_0(\sigma)$ as  
$r=\sqrt{\langle\sigma^2\rangle_{h_0(\sigma)}/
\langle\sigma\rangle^2_{h_0(\sigma)}-1}
=1/\sqrt{\nu+1}$.  
The moments of $h(\sigma)$ can be calculated as  
\begin{equation}
\langle \sigma^{\alpha}\rangle_{h(\sigma)}\equiv \int_0^{\infty}
d\sigma \sigma^{\alpha}h(\sigma)=
\frac{\Gamma(\nu+\alpha+1)}{(\nu+1)^{\alpha}\Gamma(\nu+1)}
\left(x\sigma_1^{\alpha}+(1-x)\sigma_2^{\alpha}\right).
\end{equation}

Near and above the F-S bifurcation point the density profile of 
a polydisperse mixture can be approximated as $\rho({\bf r},\sigma)
\approx\rho_0(\sigma)+\epsilon({\bf r},\sigma)$, where $\rho_0(\sigma)=
\rho_0 h(\sigma)$ is the density distribution corresponding to the 
uniform parent phase (with $\rho_0$ its number density) 
at bifurcation and $\epsilon({\bf r},\sigma)$ is 
a small non-uniform perturbation. The minimization of the grand 
potential 
\begin{eqnarray}
\Omega[\rho]&=&{\cal F}_{\rm id}[\rho]+{\cal F}_{\rm ex}[\rho]
-\int d{\bf r}\int d\sigma 
\mu_0(\sigma)\rho({\bf r},\sigma),\\
\beta {\cal F}_{\rm{id}}[\rho]&=&\int d{\bf r}\int d\sigma 
\rho({\bf r},\sigma)\left\{\ln\bigl(\mathcal{V}(\sigma)\rho({\bf r},\sigma)\bigr)-1\right\}, 
\end{eqnarray}
(where ${\cal F}_{\rm id}[\rho]$ is the ideal part of the DF while 
$\mu_0(\sigma)$ is the fixed chemical potential of species with 
length $\sigma$ and $\mathcal{V}(\sigma)$ is the thermal volume of
species with length $\sigma$) 
with respect to $\rho({\bf r},\sigma)$ gives us 
\begin{equation}
\rho({\bf r},\sigma)=
\rho_0(\sigma)\exp\left\{-\frac{\delta \beta{\cal F}_{\rm ex}}
{\delta \rho({\bf r},\sigma)}+\beta\mu_0^\mathrm{ex}(\sigma)\right\},
\end{equation}
$\beta\mu_0^\mathrm{ex}(\sigma)\equiv \beta\mu_0(\sigma)-\ln\bigl(\mathcal{V}(\sigma)\rho_0(\sigma)\bigr)$
being the excess chemical potential of species
with length $\sigma$.
Near the bifurcation point, expanding the exponential around 
$\rho_0(\sigma)$ we obtain an integral equation for the perturbation 
$\epsilon({\bf r},\sigma)$  
\begin{equation}
\epsilon({\bf r},\sigma)-\rho_0(\sigma)\int d{\bf r}'\int d\sigma' 
c^{(n)}({\bf r},\sigma,\sigma')\epsilon({\bf r}',\sigma')=0,
\label{perturba}
\end{equation}
where $c^{(n)}({\bf r},\sigma,\sigma')$ is the direct correlation 
function of the polydisperse mixture.  
The Fourier transform of Eq. 
(\ref{perturba}) for the wave vector ${\bf q}=q(1,\dots,0)$ give us 
\begin{equation}
\hat{\epsilon}(q,\sigma)-\rho_0h(\sigma)\int d\sigma'
\hat{c}^{(n)}(q,\sigma,\sigma')\hat{\epsilon}(q,\sigma')=0.
\label{dd}
\end{equation}
The function $\hat{c}^{(n)}(q,\sigma,\sigma')$ can be obtained from (\ref{note}) 
by replacing $\sigma_i$ and $\sigma_j$ by $\sigma$ and $\sigma'$ 
respectively.
We define the functions
\begin{equation}
\epsilon_{i}^{(k)}(q)=\int d\sigma \sigma^i u_k(q\sigma/2)\hat{\epsilon}
(q,\sigma),\;\; k=0,1,
\end{equation}
where we have used the notation $u_0(q\sigma/2)=\cos(q\sigma/2)$, 
and $u_1(q\sigma/2)=2\sin(q\sigma/2)/q$. Inserting 
$\hat{c}^{(n)}(q,\sigma,\sigma')$ from equation
(\ref{note}) into equation (\ref{dd}), integrating over $\sigma'$, 
multiplying the result by $\sigma^l u_k(q\sigma/2)$ 
($k=0,1$) and integrating  again over $\sigma$ we obtain
the following linear system with respect to  
the functions $\epsilon_l^{(k)}(q)$:

\begin{equation}
\epsilon_l^{(k)}(q)+
\rho_0\sum_{i=0}^{n-1}\left({\cal T}_{li}^{(k,0)}(q)
\epsilon_i^{(0)}(q)+
{\cal T}_{li}^{(k,1)}(q)\epsilon_i^{(1)}(q)\right)
=0,
\label{system}
\end{equation}
where we have defined the matrix coefficients
\begin{eqnarray}
{\cal T}^{(k,0)}_{li}(q)&=&
C_i^{n-1}\sum_{j=0}^iC_j^i\zeta_jm_{n+j+l-i-1}^{(k,1)}(q),\\
{\cal T}^{(k,1)}_{li}(q)&=&
C_i^{n-1}\sum_{j=0}^iC_j^i\left[\zeta_jm_{n+j+l-i-1}^{(k,0)}
(q)+\zeta_{j+1}m_{n+j+l-i-1}^{(k,1)}(q)\right].
\end{eqnarray}
and the generalized moments
\begin{equation}
m^{(l,n)}_i(q)=\int d\sigma h(\sigma)\sigma^i 
u_l(q\sigma/2)u_n(q\sigma/2).
\end{equation}
The functions $\zeta_i(\xi_n,\dots,\xi_{n-i})$ are obtained 
from the same recurrence relations (\ref{recurre}) with 
$\xi_i=\rho_0\langle \sigma^i\rangle$ defined now through the 
moments of the distribution functions.

The system (\ref{system}) can be put in the matrix form 
\begin{equation}
{\cal M}(\rho_0,q)
\left(
\begin{array}{cc}
\bepsilon^{(0)}(q) \\
\bepsilon^{(1)}(q)
\end{array}
\right)
={\bf 0},
\label{trivial}
\end{equation}
with 
\begin{equation}
\fl
{\cal M}(\rho_0,q)=I+\rho_0
\left(
\begin{array}{cc}
{\cal T}^{(0,0)}(q) & {\cal T}^{(0,1)}(q)\\
{\cal T}^{(1,0)}(q) & {\cal T}^{(1,1)}(q)
\end{array}
\right), \quad
\bepsilon^{(k)}=
\left(\epsilon^{(k)}_0,\dots,\epsilon^{(k)}_{n-1}\right)^T,
\end{equation}
and $I$ the $2n\times 2n$ unitary matrix. The existence of a nontrivial 
solution of Eq. (\ref{trivial}) give us the following set of equations
\begin{equation}
M(\rho_0,q)=\frac{\partial M}{\partial q}(\rho_0,q)=0,
\end{equation}
(where we have defined $M(\rho_0,q)=\det\left[{\cal M}(\rho_0,q)
\right]$) 
to find the values of $\rho_0^*$ and $q^*$ (the absolute minimum 
of $M(\rho_0,q)$ with respect to $q$) at bifurcation.

\end{document}